\documentclass[aps,prl,twocolumn,showpacs,superscriptaddress]{revtex4-1}
\usepackage{graphicx}
\usepackage{amsmath}
\usepackage{amssymb}

\begin{document}


\title{Star polymers as unit cells for coarse-graining  cross-linked networks}

\author{ Salim R. Maduar}
\affiliation{A.N.~Frumkin Institute of Physical Chemistry and Electrochemistry, Russian Academy of Sciences, 31 Leninsky Prospect, 119071 Moscow, Russia}

\author{Taras Y. Molotilin}
\affiliation{A.N.~Frumkin Institute of Physical Chemistry and Electrochemistry, Russian Academy of Sciences, 31 Leninsky Prospect, 119071 Moscow, Russia}

\author{Olga I. Vinogradova}
\email[Corresponding author: ]{oivinograd@yahoo.com}
\affiliation{A.N.~Frumkin Institute of Physical Chemistry and Electrochemistry, Russian Academy of Sciences, 31 Leninsky Prospect, 119071 Moscow, Russia}
\affiliation{Department of Physics, M.V.~Lomonosov Moscow State University, 119991 Moscow, Russia}
\affiliation{DWI - Leibniz Institute for Interactive Materials,  RWTH Aachen, Forckenbeckstr. 50, 52056 Aachen, Germany}

\date{\today}

\begin{abstract}

Reducing the complexity of cross-linked polymer networks by preserving their main  macroscale properties,  is key to understanding them, and a crucial issue is to  relate individual properties of the polymer constituents to those of the reduced network. Here we study polymer networks in a good solvent, by considering star polymers as their unit elements, and first quantify the interaction between their centers of masses. We then reduce the complexity of a network by replacing sets of its bridged star polymers  by equivalent effective soft particles with dense cores. Our coarse graining allows us to approximate complex polymer networks by much simpler ones, keeping their relevant mechanical properties, as illustrated in computer experiments on an isotropic compression.

\end{abstract}

\maketitle

One of the most difficult hurdles in the computer investigation of cross-linked polymer  networks, i.e. gels,  is, understandably, the prohibitively long  simulation time due to a large amount of particles (monomers) in the system. Explicit simulations are currently limited by nanogels with low degree of polymerization ($N\leq15$)~\cite{Claudio2009,Rumyantsev2015}. Larger systems, even microgels, represent a challenge or become  impossible to deal with. A promising way to attack this problem is to coarse-grain the complex network, i.e. to reduce the number of particles, by mapping it into a simpler one, by preserving macroscale properties. However,
general principles of such a coarse-graining have not yet been established. Some of the existing methods are currently capable to simulate only a part of a network in a periodic box~\cite{Mann2005,mann2011hydrogels,gavrilov2014computer,zidek2014mechanical}. Others do not relate the collective response of large-scale networks to individual properties of their polymer constituents~\cite{Masoud2012}.

The cross-linked network represents an aggregate of low-branched star polymers
 (SPs) connected by bridges, so that the pair interaction of network SPs could potentially be used
to construct a coarse-graining scheme.
The interaction between SPs has been studied by several groups. Most of this work focused on highly branched stars with large `dense' central core. It has been found that at short separations the potential of mean force shows a logarithmic decay and scales with functionality as $f^{3/2}$~\cite{Witten1986}. Later, a potential of mean force between two SPs that combines a short range logarithmic repulsion with a soft Yukawa-type tail has been proposed and  extensively tested~\cite{likos1998star}. The body of work investigating low branched SPs remains rather scarce. A few authors made important remarks that properties of low branched SPs  could differ from those of highly branched~\cite{jusufi2001effective, rai2016, chen2016}. Thus, when $f\leq 10$  the monomer density around the central bead is no longer described by the blob model, and the Yukawa-type repulsion is not observed~\cite{jusufi2001effective,rubio2000interaction}. We should note that contrary to interacting linear chains ($f=2$), where coordinates associated with the centers of mass are normally employed~\cite{louis2000can,louis2000mean,Bolhuis2001,kruger1989correlations}, previous studies of SPs have commonly used coordinates related to central beads~\cite{likos1998star,jusufi2001effective}. Although some investigations of low branched stars~\cite{rubio2000interaction,rubio1996monte} have used the center of mass approach, they did not attempt to predict star interaction energies. Finally, we note that prior work concerned an interaction of SPs in solutions only and not attempted to calculate it for SPs, constituting the network.

\begin{figure}[t]
	\centering
	\includegraphics[scale=0.16]{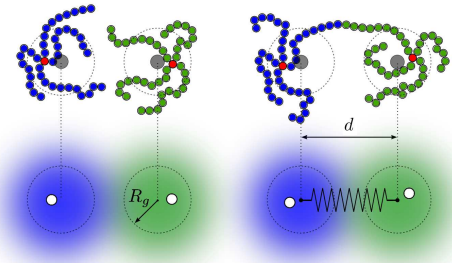}
	\caption{\label{fig:chain-model} Sketch of interacting star polymers of $R_g$ separated by distance $d$. Explicit bead-spring models are shown at the top, and coarse-grained models - at the bottom. Red, gray and white circles indicate central beads, centers of mass, and `dense' cores. }
\end{figure}

\begin{figure}[t]
	\centering
	\includegraphics[scale=0.10]{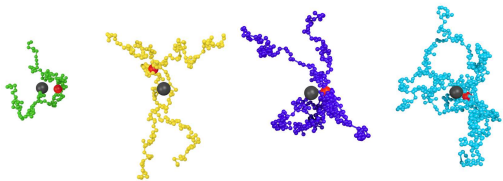}
	\caption{\label{fig:star} Simulation snapshots of star polymers with $N=33$ and $f=2, 4, 6, 8$. Large gray beads indicate the centers of mass. Red beads are the central cores.}
\end{figure}

In this Letter we propose a procedure to coarse-grain polymer networks in a good solvent, based on the idea of mapping  their low branched SPs to effective particles.  Our starting point is the analysis of interactions between two network SPs, by using centers of mass as effective coordinates  (see Fig.~\ref{fig:chain-model}). We obtain  expressions for potentials of mean force, which define a coarse-grained model, and validate them by explicit (monomer-resolved) simulations~\cite{epaps}. Finally, we illustrate the power of our approach by studying mechanical deformations of the reduced  networks, and show that coarse-graining provides a highly representative approximation of the initial network, by dramatically reducing an amount of particles in simulations.


We consider theoretically two interacting identical SPs of fixed (to arbitrary values) $f$ and $N$ in a good solvent. Although the SPs are normally defined for $f\geq 3$, we also consider a  special case of $f=2$, which corresponds to a linear polymer chain of a degree of polymerisation $2N$.

\begin{figure}[t]
\includegraphics[width=0.95\columnwidth]{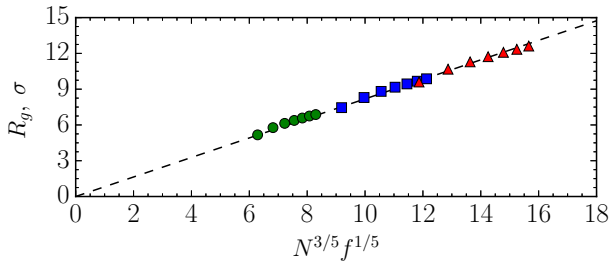}
	\caption{\label{fig:gyration} Radii of gyration obtained in MC simulations for star polymers of $N=17$ (circles), $32$ (squares), and $49$ (triangles). Symbols from left to right show data for $f$ from $2$ to $8$. Dashed line corresponds to the scaling relationship $R_g\propto N^{3/5}f^{1/5}$. 
	}
\end{figure}

\begin{figure}[t]
	\centering
\includegraphics[width=0.95\columnwidth]{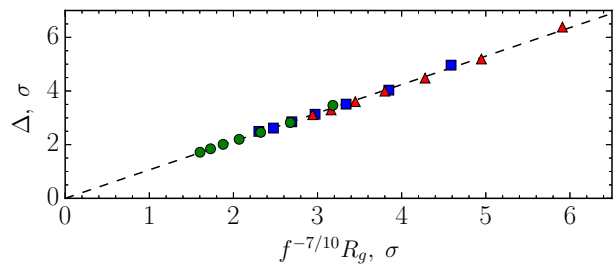}
	\caption{\label{fig:cm-cb} Root mean-squared distance between the center of mass and the central bead of a star polymer of functionality $f = 2-8$ obtained in MC simulations. Circles show simulation data for $N=17$, squares - for $N=32$  and triangles - for $N=49$. Dashed line plots predictions of Eq.(\ref{eq:cm-cb}). }
\end{figure}

We first justify the choice of our effective coordinates. Note that the time average location of the center of mass of an SP does of course coincide with that of the central bead, but its instantaneous position deviates from the central bead location. This can be illustrated simulations of single SPs with different $f$ and $N$ (shown in Fig.~\ref{fig:star}). Interactions between monomers (of size $\sigma$) are described by Lennard-Jones potential with $\epsilon_{LJ}/k_BT = 0.05$~\cite{epaps}.  The simulation data shows discernible deviations of a central bead from the center of mass, which, however, tend to decrease with $f$. A corollary from this is that the mean-squared distance $\Delta^2$ between the center of mass and the core is finite. Indeed, one can show using mean-field arguments~\cite{epaps} that $\Delta$ scales with $f$ as
\begin{equation}
	\label{eq:cm-cb}
	\Delta = A  f^{-7/10} R_g,
\end{equation}
which indicates that $\Delta$ could be comparable with the radii of gyration, $R_g$. To  prove this we have first measured $R_g$ in simulations and plotted it in Fig.~\ref{fig:gyration}. One can see that our data agrees well with the scaling law, $R_g \propto N^{\nu}f^{1/5}$ (with the Flory exponent $\nu\simeq3/5$), suggested earlier for SPs in a good solvent~\cite{Daoud1982}. We have then obtained the values of $\Delta$ (see Fig.~\ref{fig:cm-cb}), and fitted the simulation data to Eq.(\ref{eq:cm-cb}) taking $A$ as
a fitting parameter. The theoretical curve is included in Fig.~\ref{fig:cm-cb} and the value $A=1.07$ has been obtained from fitting. This is close to $A= 2^{-2/5}\sqrt{11/5} \simeq 1.12$ predicted by our mean-field theory~\cite{epaps}. These results demonstrate that the central bead fluctuates around the center of mass, which is especially pronounced at low $f$. Therefore, the commonly used central bead poorly  represents the location of the star, so that below we use the centers of mass as effective coordinates of SPs.

\begin{figure}
\centering
\includegraphics[width=0.95\columnwidth]{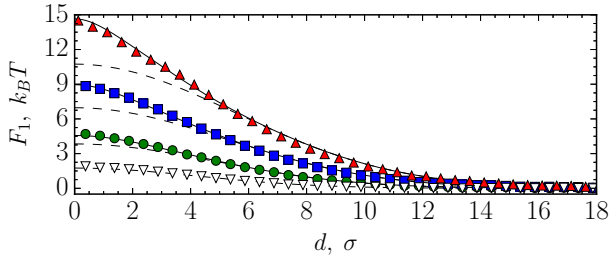}
	\caption{\label{fig:pot-full} Interaction potentials $F_1$ between star polymers of $N=17$ obtained in simulations (symbols). From top to bottom $f=8$, $6$, $4$, $2$. Dashed curves show $F^{ss}$ predicted by Eq.(\ref{eq:soft-sphere}),  solid curves are fits to Eq.(\ref{eq:pot-full}).}
\end{figure}

Let us now investigate the effect of functionality on the value of the  interaction free energy $F_1$ of two SPs as a function of separation $d$ between their centers of mass.  This has been calculated by using the histogram method with a bias potential to ensure efficient sampling of configuration space~\cite{Frenkel:2001,ferrenberg1989optimized} as described in~\cite{epaps}. In Fig.~\ref{fig:pot-full} we plot simulation results obtained at fixed $N=17$ and $f$ varying  from $2$ to $8$. The data show that the two SPs always repel each other, and that the value of $F_1$ increases with $f$. Remarkably, it remains finite at zero separation, i.e. when centers of mass overlap, and there is no manifestation of logarithmic divergence at $d=0$ predicted when central beads are chosen as effective coordinates~\cite{likos2001soft}. We note that the `soft' repulsion of SPs in our case resembles that of linear chains~\citep{Bolhuis2001}.


To interpret simulation data we first consider the long-range or `soft sphere' part of interaction, which is  attributed to SP's coronas. In the case of linear chains, $f=2$, the interaction free energy  can be  represented by a Gaussian function,  $F_1 = F^{ss} = F_0 \exp\left(-3d^2/4R_g^2\right)$~\cite{Flory1950}, with a range of the order of $R_g$~\cite{louis2000mean,Bolhuis2001} and $F_0 \simeq 1.53 k_B T$ found theoretically~\cite{kruger1989correlations}. Note, however, that some simulations have deduced $F_0 \simeq 1.9 k_B T$~\citep{Bolhuis2001,louis2000can}. The scaling expression for $F_0$ in case of SPs can be estimated using average number of  contacts between monomers, corrected for their correlations, as~\cite{daoud1975,Grosberg1982}
\begin{equation}
	\label{eq:overlap0}
\frac{F_0}{k_BT} \propto \rho V \left(\rho a^3\right)^{1/\left(3\nu-1\right)}\simeq\frac{f^{9/4}N^{9/4}}{R_g^{15/4}},
\end{equation}
where we have used  $V \simeq R_g^3$ for the overlap volume and $\rho \propto {fN}/{R_g}$ for monomer density. By substituting scaling expressions for  $R_g$ into Eq.\eqref{eq:overlap0} we obtain $F_0 \propto f^{3/2}$ and the `soft sphere' interaction free energy becomes similar to known for interacting linear polymers:
\begin{equation}
	\label{eq:soft-sphere}
		F^{ss} = F_0\exp\left(-\frac{3d^2}{4R_g^2}\right),
\end{equation}
but includes $F_0$, which depends on $f$.  Note, however, that $F_0$ does not depend on $N$, which is similar to results for linear polymers~\cite{Grosberg1982,Bolhuis2001}.

\begin{figure}[t]
	\centering
\includegraphics[width=0.95\columnwidth]{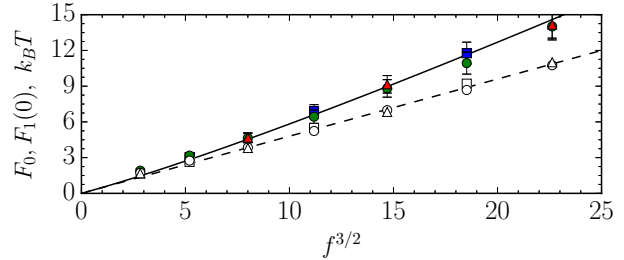}
	\caption{\label{fig:scaling-n1n2} Amplitude of the `soft sphere' (open symbols) and total (closed symbols) potential as a function of $f^{3/2}$ for $N=17$ (circles), $25$ (squares), and $49$ (triangles). Solid and dashed lines are added to guide the eye. }
\end{figure}

Calculations made using Eq.\eqref{eq:soft-sphere} with $F_0$ taken as adjustable parameters for the long-range tails are included in Fig.~\ref{fig:pot-full}. We see that simulation data at large $d$ are indeed well described by a Gaussian repulsion with $R_g$ found above (see Fig.\ref{fig:gyration}).  To verify the scaling relationship for $F_0=F^{ss} (0)$ we now plot it in Fig.~\ref{fig:scaling-n1n2} as a function of $f^{3/2}$. Also included are additional simulation data for $N = 25$ and $49$.  These data allows us to deduce the universal value of $F_0=(0.48\pm0.03)~f^{3/2}k_BT$, which is valid for all $f$ and does not depend on $N$. Fig.~\ref{fig:scaling-n1n2} also includes the SPs interaction free energy, $F_1$, obtained from simulation data at $d=0$. One can conclude that  for all $N$ deviations of $F (0)$ from $F_0$ are negligibly small when $f=2$ and $3$, but they become discernible at larger functionalities, and their values increase with $f$. The discrepancy is always in the direction of larger potential than $F_0$.

We remark that deviations of $F_1$ from $F^{ss}$ given by Eq.(\ref{eq:soft-sphere}) are seen only at small $d$ as seen yet in Fig.~\ref{fig:pot-full}.
An explanation can be obtained if we invoke the short-range repulsion emerging when SPs strongly overlap, so that the  effective interaction between their `dense' cores becomes important. For simplicity we model the entropic short-range logarithmic interaction of the cores~\cite{Witten1986} by describing them as `hard-spheres' of an effective radius $\sigma_c$.
Then each `dense' core may be seen as a Brownian particle of diffusion coefficient $D$ fluctuating around the center of mass with zero mean, but finite variance, $\Delta^2$, see Eq.(\ref{fig:cm-cb}).  The interaction free energy is then given by $F^{hs}=-k_BT \ln (1-P)$, where $P(t, d)$ is the probability for a collision of two dense cores, initially separated by distance $d$, to occur after time $t$ given by $6Dt = \Delta^2$. Thereby in $F^{hs}$ we exclude configurations where cores  approach closer than $2\sigma_c$. The solution for $P$ may be found by considering properties of diffusing particles~\cite{epaps}:

\begin{equation}\label{eq:hs-prob}
P = \dfrac12 {\rm{erf}}(y_-)-\dfrac12 {\rm{erf}}(y_+) + \sqrt{\dfrac{\Delta^2}{3 \pi d^2}} \left(e^{-y_+^2}-e^{-y_-^2}\right),
\end{equation}
where $y_{\pm} = \displaystyle \sqrt{3}\dfrac{d \pm 2\sigma_c}{2\Delta }$. For  small $f$ or for $\sigma_c/\Delta\ll1$  the interaction free energy of `dense' cores reduces to a Gaussian function:

\begin{equation}
F^{hs} = k_BT  \dfrac{4\sqrt{3}}{\sqrt\pi} {\exp\left(-{\frac{3d^2}{4 \Delta^2}}\right)} \left(\dfrac{\sigma_c}{\Delta}\right)^3
\label{eq:hard-sphere}
\end{equation}
We remark that although the cores are represented by `hard-spheres', their interaction free energy may still be finite at $d=0$.
In our simulations we found that $\sigma_c/\Delta \simeq B f$, so it is independent on $N$. Here $B$ is constant for all $f,N$ which was found to be equal to $\simeq0.16\pm0.02$. This implies that $\sigma_c$ scales as $f^{1/2}$, which is in agreement with prior work~\cite{Daoud1982}.
We also note that with our parameters for $f=2$ we have $\sigma_c/\Delta \simeq 0.3$, so that at $d=0$ this gives $F^{hs} \simeq 0.1 k_B T$, which is much smaller than $F_0$ and can safely be neglected. Eq.(\ref{eq:hard-sphere}) can be used to describe SPs up to $f=4$. Finally, in the limit of large $f$ our Eq.(\ref{eq:hs-prob}) reduces to the `hard-sphere' interaction  potential. We should like to stress that unlike logarithmic repulsion,  $F^{hs}$ vanishes at large $d$, so that we do not need to adjust the cut-off distance for a short-range interaction as it has been done before~\cite{likos1998star}.

Now combining both soft-sphere and hard-sphere repulsions we can propose the repulsive potential of mean force for two SPs

\begin{equation}
	\label{eq:pot-full}
	F_1 = F^{ss}+F^{hs}
\end{equation}
with $F^{ss}$ and $F^{hs}$ defined by Eqs.(\ref{eq:soft-sphere}) and (\ref{eq:hard-sphere}). Theoretical curves calculated with Eq.(\ref{eq:pot-full}) are included in Fig.~\ref{fig:pot-full}. We see that our model is in excellent agreement with simulation data for all $d$.


We finally turn to two SPs as a network segment. The important difference from the solutions 
 is the bridging of SPs, which should give  rise to an additional attraction between them. This bridging attraction, $F^b$,  should be added to Eq.\eqref{eq:pot-full} to give
\begin{equation}\label{eq: bonded}
    F_2 =F^{ss}+F^{hs}+F^{b}.
\end{equation}
As long as $d\ll 2 N \sigma$, $F^{b}$ can be can be estimated as the free energy of stretching of a linear chain
\begin{equation}
	\label{eq:bonded-full}
	F^{b} = \frac{kd^2}{2} ,
\end{equation}
where $k= 3 k_BT/R_F^{2} $ with $R_F \simeq (2N)^\nu\sigma$~\cite{deGennes1979,FloryBook1953}, but note that for very large $d\gg R_g$ one has to define $F^b$ differently~\cite{pincus1976excluded,grosberg1994statistical}. We also stress that since the bridging attraction is long-range, $d\gg \Delta$. Therefore, this contribution does not depend on the choice of coordinates.

\begin{figure}[t]
	\centering
	\includegraphics[width=0.95\columnwidth]{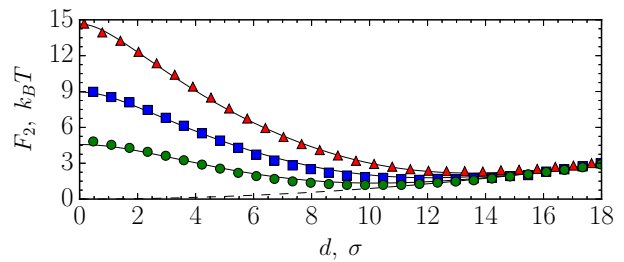}
	\caption{\label{fig:pot-full-spring} Interaction potential of two SPs connected by a bridge of $2N=34$ (symbols). From top to bottom $f=8, 6,$ and $4$. Solid curve shows calculations with Eq.(\ref{eq: bonded}). Dashed curves plots bridging attraction given by Eq.(\ref{eq:bonded-full}).}
\end{figure}

To verify the model we have simulated the potentials of mean force between two SPs of $f$ varying from $2$ to $8$ connected via a bridge of fixed $2N=34$. The values of $F$ obtained in simulations are plotted in Fig.~\ref{fig:pot-full-spring}. This plot also includes theoretical curves calculated with Eq.(\ref{eq: bonded}). The calculations are made using $R_g$ shown in Fig.~\ref{fig:gyration}  and  the ratio $R_g^2/R_F^2 \simeq 0.157$~\cite{wall1959statistical} leading to $k \simeq 1.22 N^{-6/5} k_B T/\sigma^2$. In other words, there are no adjustable parameters in the theoretical curves. We see that the fits are very good for all $d$, which confirms the validity of our model.
Another important conclusion from Fig.~\ref{fig:pot-full-spring} is that $F_2$ has a minimum at $d_0 \simeq 2R_g\sqrt{\ln (3F_0/2kR_g^2)} /\sqrt3$, which corresponds to the equilibrium position of two SPs. Therefore, they may be seen as an effective spring of a constant $k_{\rm eff} \simeq 2k\ln (3F_0/2kR_g^2)$. To verify the model for $k$ we have made simulations for SPs of $f=2$ and $4$, and $N$ varying from $17$ to $49$, and found that results fully confirm our theory~\cite{epaps}. Alltogether these suggest that a polymer network segments (SPs) can be effectively represented by soft Gaussian spheres with `hard' cores connected by springs.

\begin{figure}[t]
	\centering
	\includegraphics[width=0.95\columnwidth]{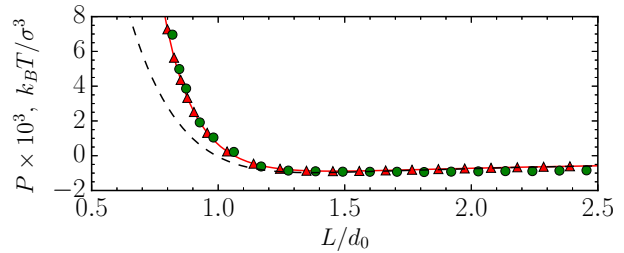}
	\caption{Pressure as a function of $L/d_0$ for a network composed of SPs of $N=17$ and $f=6$ with $d_0 \simeq 12.1\sigma$.  \label{fig:pot-gel} }
\end{figure}

Finally, we perform explicit MD simulations of a deformed cross-linked network with an open-source package ESPResSo~\cite{espresso}. Specifically, we study a primitive cubic network of SPs of $f=6$ connected by bridges of $2N =34$, and measure a pressure, $P$, as a function of the size of the unit cell $L$. We also perform the coarse-grained simulations, where we replace the network SPs by effective spheres interacting with each other with potentials $F_1$ and $F_2$, which reduces the number of particles in $f N$ times and therefore significantly accelerates calculations. The detailed comparison between the explicit simulation results and the coarse-graining approach is then shown in Fig.~\ref{fig:pot-gel}. A general conclusion from this plot is that the coarse-graining data are in excellent agreement with explicit simulation results. Note that one can also roughly evaluate pressure theoretically as $P = -\dfrac{f  }{2} dF_2/dV$, i.e. by neglecting interactions of SPs, which are not connected by bridges. Here $f/2$ is the number of bridges in volume $V=L^3$. These estimates are also included in Fig.~\ref{fig:pot-gel}, and show that this simple theory agrees well with simulation data for $L/d_0 = O(1)$ and larger, i.e. for stretching. However, $F_1$ cannot be ignored in the case of compression, i.e. small $L/d_0$.

In conclusion, we have calculated the free interaction energy of two identical SPs by using their centers of mass as effective coordinates. Our analysis has led to explicit expressions for interaction potentials of SPs of any $f$ and $N$, and in the limiting case of $f=2$ recovers known results for linear polymers. We have checked the validity of our theory by explicit MC simulations. These potentials have provided a framework for a coarse-graining approach, allowing one to reduce the number of particles in simulations in $f N$ times without losses in accuracy. The advantages of our coarse-graining method have been illustrated by considering a compression of an ideal polymer network, but our results can of course immediately be applied to studying various mechanical properties of non-ideal polydisperse networks, or to a situation, where entanglements become important.

We thank O.V.Borisov and F.Schmid for helpful discussions, and K.Binder for valuable comments on the manuscript. The simulations were carried out using computational resources at the Moscow State University (`Lomonosov' and `Chebyshev').

\bibliography{microgel}

\end{document}